\documentclass[prl,twocolumn,superscriptaddress,showpacs,amsmath,amssymb]{revtex4}
\usepackage{graphicx}
\DeclareMathOperator{\trace}{Tr}

\newcommand\rmd{\, \mathrm{d}}
\newcommand\rme{\mathrm{e}}
\newcommand\rmi{\mathrm{i}}
\newcommand\G{\mathrm{G}}
\newcommand\bA{\overline{A}}
\newcommand\bB{\overline{B}}
\newcommand{\abs}[1]{\lvert #1 \rvert}
\newcommand{\ket}[1]{\left | #1 \right \rangle}
\newcommand{\bra}[1]{\left \langle #1 \right |}

\newcommand{\UN}{\mathrm{U}(N)}
\newcommand{\ONen}{\mathrm{O}^{-}(2N +2)}
\newcommand{\ONep}{\mathrm{O}^{+}(2N)}
\newcommand{\ONo}[1]{\mathrm{O}^{#1}(2N + 1)}

\newcommand{\SptN}{\mathrm{Sp}(2N)}

\newcommand{\vphi}{\boldsymbol{\phi}}
\newcommand{\vpsi}{\boldsymbol{\psi}}
\newcommand{\gsk}{\ket{\boldsymbol{\Psi}_{{\rm g}}}}
\newcommand{\gsb}{\bra{\boldsymbol{\Psi}_{{\rm g}}}}

\begin{document}
\title{Entanglement in Quantum Spin Chains, Symmetry Classes of Random Matrices, and Conformal Field Theory}
\author{J.P.\ Keating}%
\author{F.\ Mezzadri}
\affiliation{School of Mathematics, University of Bristol,
Bristol BS8 1TW, UK.}%
\begin{abstract}
We compute the entropy of entanglement between the first $N$ spins
and the rest of the system in the ground states of a general class
of quantum spin-chains. We show that under certain conditions the
entropy can be expressed in terms of averages over ensembles of
random matrices. These averages can be evaluated, allowing us to
prove that at critical points the entropy grows like $\kappa\log_2
N + {\tilde \kappa}$ as $N\rightarrow\infty$, where $\kappa$ and
${\tilde \kappa}$ are determined explicitly. In an important class
of systems, $\kappa$ is equal to one-third of the central charge
of an associated Virasoro algebra. Our expression for $\kappa$
therefore provides an explicit formula for the central charge.
\end{abstract}
 \maketitle

Entanglement has recently come to be viewed as an important
physical resource for manipulating quantum information. The
problem of quantifying it is, however, still poorly understood,
especially when the entanglement is shared between more than two
systems. When the entanglement of a pure state is shared between
two parties, i.e.~in a {\it bipartite system}, Bennett {\em et
al}~\cite{BBPS96} have shown that it is consistent to define it as
the von Neumann entropy of either of the two parts. We consider
here the general class of quantum spin chains arising from
quadratic chains of fermionic operators in their ground state.
These systems are partitioned into two contiguous subchains. If
the ground state is non-degenerate, this subdivision creates a
pure bipartite system; our main result is to calculate its
entanglement entropy by relating the problem to one in random
matrix theory.

As is well known, the systems we are studying exhibit quantum
phase transitions.  These manifest themselves as qualitative
changes in the decay of correlations: algebraic at a critical
point and exponential decay away from it. Entanglement plays a
fundamental role in the quantum phase transitions that occur in
interacting lattice systems at zero
temperature~\cite{AOPFP04,VLRK02,JK03,Kor04,CC04,LEB04,VMC04}.
Under these conditions the system is in the ground state, which is
also a pure state, and any correlations must be a consequence of
the fact the ground state is entangled. It follows immediately
that the entanglement changes qualitatively at critical points.

In this context, Vidal {\it et al.}~\cite{VLRK02} studied the
ground states of a wide range of one-dimensional spin models
partitioned into two consecutive subchains.  They observed
numerically that, when the Hamiltonian undergoes a phase
transition, the entanglement of formation of these bipartite
systems grows logarithmically with the size $N$ of one of the two
parts.  Jin and Korepin~\cite{JK03} then proved that the entropy
grows like $\frac{1}{3}\log_2 N$ in the XX model, for which the
Hamiltonian is
\begin{equation}
\label{XYmodel} H_{\alpha} = -\frac{\alpha}{2}\sum_{j = 0}^{M-1}
\left[\sigma_j^x \sigma_{j+1}^x + \sigma_j^y \sigma_{j+1}^y\right]
- \sum_{j=0}^{M-1} \sigma_j^z,
\end{equation}
where $\sigma^a$ denotes the Pauli matrices and $a=x,y,z$.
Recently, Korepin~\cite{Kor04} and Calabrese and Cardy~\cite{CC04}
showed, using conformal field-theoretic arguments developed by
Holzhey {\it et al.}~\cite{HLW94}, that the logarithmic divergence
of the entanglement in one dimensional systems is a general
consequence of the logarithmic growth of the entropy with the size
of the system at phase transitions. These arguments determine the
constant multiplying the leading order $\log_2 N$ term in the
asymptotics to be one-third of the central charge of the
associated Virasoro algebra~\cite{Vira}.

We show here that if a quantum spin-chain Hamiltonian posses
certain symmetries, the entanglement can be expressed as an
average over an ensemble of random matrices corresponding to one
of the classical compact groups equipped with Haar measure, i.e.
one of the following groups: $\mathrm{U}(N)$, $\mathrm{Sp}(2N)$
and $\mathrm{O}^\pm(N)$, where the superscript $\pm$ indicates the
connected component of the orthogonal group with determinant $\pm
1$. From the point of view of the entanglement entropy (and of
spin-spin correlations), quantum spin chains therefore divide into
symmetry classes related to the classical compact groups.  The XX
model turns out to be an example of a system with $\mathrm{U}(N)$
symmetry. The averages that occur can be expressed either as
Toeplitz determinants (i.e.~determinants of matrices in which the
elements are functions of the difference between the row and
column indices), in the case of $\UN$, or as determinants of
specific combinations of Toeplitz and Hankel matrices
(i.e.~matrices in which the elements are functions of the sum of
the row and column indices) for the other compact groups.
Asymptotic formulae for these determinants then lead to general
expressions for the leading-order and next-to-leading-order terms
in the asymptotics of the entanglement in the limit as the total
number of spins tends to infinity and then as
$N\rightarrow\infty$.

We find that at a critical point the entanglement grows
logarithmically with $N$, in agreement with the
conformal-field-theoretic calculations. We derive a general
formula for the associated constant of proportionality. This is a
rational number, the numerator of which is shown to factorize into
a universal part, related to symmetries of the quantum Hamiltonian
and which can be calculated from the random-matrix averages, and a
non-universal (i.e.~Hamiltonian-specific) part, which we also
evaluate. In the unitary case, comparing with the results
in~\cite{HLW94, Kor04, CC04} leads to an explicit formula for the
central charge. However, our approach also extends to systems
where the conformal-field-theoretic results cannot be applied
directly. These are the systems related to the other compact
groups. Further details of our calculations and results may be
found in~\cite{KM04}.

The most general form of Hamiltonian related to quantum spin
chains is
\begin{multline}
\label{impH}  H_{\alpha} = \alpha \left[\sum_{j,k=0}^{M-1}
b^\dagger_jA_{jk} b_k + \frac{\gamma}{2}\left(b^\dagger_j
B_{jk}b^\dagger_k - b_j B_{jk}b_k \right)\right] \\ -
2\sum_{j=0}^{M-1} b^{\dagger}_jb_j,
\end{multline}
where $\alpha$ and $\gamma$ are real parameters, $0\le \gamma \le
1$, the $b_j$s are Fermi oscillators, $A$ is an Hermitian matrix,
and $B$ is an antisymmetric matrix. We take periodic boundary
conditions, i.e. $b_M = b_0$.  Without loss of generality, we will
consider only matrices $A$ and $B$ with real elements.  The
Hamiltonian (\ref{impH}) can always be re-expressed in terms of
the Pauli spin matrices using the Jordan-Wigner
transformation~\cite{LSM61, KM04}.

We will here be concerned with the entanglement between the first
$N$ oscillators and the rest of the chain when the system is in
the ground state $\gsk$ and as the length of chain tends to
infinity. We decompose the Hilbert space into the direct product
$\mathcal{H}=\mathcal{H}_{{\rm P}} \otimes \mathcal{H}_{{\rm Q}}$,
where $\mathcal{H}_{{\rm P}}$ is generated by the first $N$
sequential oscillators and $\mathcal{H}_{{\rm Q}}$ by the
remaining $M - N$. Our goal is to determine the asymptotic
behaviour for large $N<<M$ of the von Neumann entropy $E_{{\rm P}}
= - \trace \rho_{{\rm P}} \log_2 \rho_{{\rm P}}$, where
$\rho_{{\rm P}}= \trace_{{\rm Q}} \rho_{{\rm PQ}}$ and $\rho_{{\rm
PQ}}= \gsk\negthinspace\gsb$.

The first step involves determining the expectation values with
respect to $\gsk$ of products of arbitrary numbers of the
operators
\begin{equation}
\label{mop} m_{2l + 1} = \left(\prod_{j =0}^{l-1} \sigma_j^z
\right)\sigma_l^x \quad {\rm and} \quad m_{2l} =
\left(\prod_{j=0}^{l-1} \sigma_j^z\right)\sigma_l^y.
\end{equation}
From the invariance of the Hamiltonian~\eqref{impH} under the
transformation $b_j \mapsto -b_j$, it follows that $\gsb m_l
\gsk=0$; for the same reason, the expectation value of the product
of an odd number of $m_j$s must be zero. The expectation values
$\gsb m_j m_k \gsk$ can be deduced using the approach of Lieb {\it
et al}~\cite{LSM61}: $\gsb m_jm_k \gsk  = \delta_{jk} + \rmi
(C_M)_{jk}$,
where the correlation matrix $C_M$ has the block structure
\begin{equation}
\label{cmat}
C_M = \begin{pmatrix} C_{11} & C_{12} & \cdots & C_{1M} \\
                      C_{21} & C_{22} & \cdots & C_{2M} \\
                      \hdotsfor[2]{4} \\
                      C_{M1} & C_{M2} & \cdots & C_{MM}
\end{pmatrix}
\end{equation}
with
\begin{equation}
\label{blockC}
C_{jk} = \begin{pmatrix} 0 &  (T_M)_{jk}\\
                         -(T_M)_{kj} & 0
         \end{pmatrix},
\end{equation}
the matrix $T_M$ is defined by
\begin{equation}
\label{Tmatr} \left(T_M\right)_{jk} = \sum_{l=0}^{M-1}\psi_{l
j}\phi_{lk}, \quad j,k=0,\ldots, M-1,
\end{equation}
and the vectors $\vphi_k$ and $\vpsi_k$ are real and orthogonal
and obey the eigenvalue equations
\begin{subequations}
\label{stepb}
\begin{align}
\label{phi2b} \alpha^2\left(A -\frac{2}{\alpha}I-\gamma
B\right)\left(A -\frac{2}{\alpha}I +\gamma B\right)\vphi_k & =
\abs{\Lambda_k}^2\vphi_k, \\
\label{psi2b} \alpha^2\left(A -\frac{2}{\alpha}I+\gamma B\right)
\left(A -\frac{2}{\alpha}I-\gamma B\right)\vpsi_k & =
\abs{\Lambda_k}^2\vpsi_k.
\end{align}
\end{subequations}
These vectors are related by
\begin{subequations}
\label{fstepb}
\begin{align}
\label{phi1b}
\alpha \left(A -\frac{2}{\alpha}I + \gamma B\right)\vphi_k &=
\abs{\Lambda_k} \vpsi_k, \\
\label{psi1b}
\alpha\left(A - \frac{2}{\alpha}I - \gamma
B\right)\vpsi_k &= \abs{\Lambda_k}\vphi_k.
\end{align}
\end{subequations}
The expectation values of the product of an even number of $m_j$s
can then be computed using Wick's theorem (see, for
example,~\cite{BD}).

Following the calculation in~\cite{JK03}, the formula for the
entropy of the subchain P that one obtains using these expressions
for the expectation values is then
\begin{equation}
\label{JKidea2} E_{{\rm P}} = \lim_{\epsilon \rightarrow 0^+}
\lim_{\delta \rightarrow 0^+} \frac{1}{2\pi \rmi}
\oint_{c(\epsilon,\delta)} \rme(1 + \epsilon,\lambda) \frac{\rmd
\ln D_N(\lambda)}{\rmd \lambda} \rmd \lambda,
\end{equation}
where
\begin{equation}
\label{binaryent} \rme(x,\nu)= - \frac{x +
\nu}{2}\log_2\left(\frac{x + \nu}{2}\right) - \frac{x -
\nu}{2}\log_2\left(\frac{x - \nu}{2}\right),
\end{equation}
$D_N(\lambda) = \det\left(I\lambda - S \right)$, $S$ is the real
symmetric matrix $\left(T_N T_N^t\right)^{1/2}$ and $T_N$ is
obtained from the matrix~\eqref{Tmatr} by removing the last $M-N$
rows and columns. The contour of integration $c(\epsilon,\delta)$
depends on the parameters $\epsilon$ and $\delta$ and includes the
interval $[-1,1]$; as $\epsilon$ and $\delta$ tend to zero the
contour approaches the interval $[-1,1]$.  This guarantees that
the branch points of $\rme(1 + \epsilon,\lambda)$ lie outside the
contour of integration and thus that $\rme(1 + \epsilon,\lambda)$
is analytic inside $c(\epsilon,\delta)$.  The eigenvalues of $S$
must all lie in the interval $[-1,1]$, and this is the case for
the various Hamiltonians we consider~\cite{KM04}.

We first specialize to cases where the Hamiltonian~\eqref{impH} is
invariant under translations.  For example, the XX model has this
symmetry.

We denote $\bA =\alpha A - 2I$ and $\bB = \alpha \gamma B$. If
$H_{\alpha}$ is invariant under translations of the lattice
$\{0,1,\ldots,M-1\}$, then the elements of the matrices $\bA$ and
$\bB$ must depend only on the difference between the row and
column indices, i.e. $\bA$ and $\bB$ must be Toeplitz matrices. In
addition, because of the periodic boundary conditions, $\bA$ and
$\bB$ must be {\it cyclic}.

Now, let $a$ and $b$ be two real functions on
$\mathbb{Z}/M\mathbb{Z}$, even and odd respectively. The matrix
elements of $\bA$ and $\bB$ can be written as
\begin{equation}
\label{circdef} \bA_{jk}=a(j-k) \quad {\rm and} \quad
\bB_{jk}=b(j-k).
\end{equation}
The complex exponentials
\begin{equation}
\label{ceigAB} \phi_{kj}= \frac{\exp\left(\frac{2\pi \rmi k
j}{M}\right)}{\sqrt{M}}, \quad j,k=0,\ldots, M-1,
\end{equation}
form a complete orthonormal set of eigenvectors of cyclic
matrices, as can be easily verified by direct substitution. The
matrices $\bA$ and $\bB$ defined in~\eqref{circdef} commute. As a
consequence, the complex exponentials~\eqref{ceigAB} are a
complete set of eigenvectors of $\bA + \bB$ too.

The eigenvalues of $\bA + \bB$ can be determined by inserting the
eigenvectors~\eqref{ceigAB} into the eigenvalue equations and
using the parities of the functions $a(j)$ and $b(j)$. We have
that when $M$ is odd
\begin{equation}
\label{fexpu1} \Lambda_k =  a(0) + 2\sum_{j =
1}^{(M-1)/2}\left[a(j)\cos kj + \rmi b(j)\sin kj\right]
\end{equation}
and when $M$ is even
\begin{equation}
\begin{split}
\label{fexpu2} \Lambda_k & =a(0) + (-1)^l a(M/2)  \\ & \quad +
2\sum_{j = 1}^{M/2-1}\left[a(j)\cos kj  + \rmi b(j) \sin
kj\right],
\end{split}
\end{equation}
where $k = 2\pi l/M$.  Then
\begin{equation}
\label{Toeplitzr} (T_N)_{jk} \xrightarrow[M \rightarrow \infty]{}
 \frac{1}{2\pi}\int_0^{2\pi}\frac{\Lambda(\theta)}{\abs{\Lambda(\theta)}}
\rme^{-\rmi \left(j - k\right)\theta}\rmd \theta,
\end{equation}
where $\Lambda(\theta)$ is the periodic function
\begin{equation}
\label{fLam} \Lambda(\theta)=\sum_{j=-\infty}^\infty \Lambda_j \,
\rme^{\rmi j \theta}
\end{equation}
with $\Lambda_j=a(j)-b(j)$ if $j>0$ and $\Lambda_j=a(j)+b(j)$ if
$j<0$. Note that $T_N[g]$ is a Toeplitz matrix. $g(\theta)=
\Lambda(\theta)/\abs{\Lambda(\theta)}$ is called the symbol of
$T$. It is worth emphasizing that~\eqref{Toeplitzr} has been
obtained by assuming only the translation invariance of the
Hamiltonian~\eqref{impH} and periodic boundary conditions.

We now make the key observation that when $T_N[g]$ is symmetric
the matrix $S$, which appears in the definition of $D_N(\lambda)$,
is equal to $T_N$. We can then apply a famous identity of
Heine~\cite{Hei78} and Szeg\H{o}~\cite{Seg59}, which asserts that
if $G(U)$ is a function on $\UN$ that depends only on the
eigenvalues $\exp (i\theta_j)$ of $U$ and is such that $G(U) =
\prod_{j=1}^{N}g(\theta_j)$, where $g(\theta)$ is $2\pi$-periodic,
then
\begin{equation}
\label{HSid} \Bigl \langle G(U) \Bigr \rangle_{{\UN}} =
\det\left(g_{j-k}\right)_{j,k=0,\ldots,N-1},
\end{equation}
where $g_l$ is the $l$th Fourier coefficient of $g$.  In our
context, this implies that~\eqref{JKidea2} can be expressed as an
average with respect to Haar measure over the unitary group $\UN$,
i.e.~over the Circular Unitary Ensemble of $N\times N$ random
matrices. A necessary and sufficient condition for $T_N[g]$ to be
symmetric is that $\Lambda(\theta)$ should be real and even, or
equivalently $\gamma$ should be zero; in other words, the
interaction in the Hamiltonian~\eqref{impH} must be isotropic.
When $\gamma =0$ the symbol $g(\theta)$ is a piece-wise continuous
function which takes the values $1$ and $-1$ and has
discontinuities at all points $\theta_r$ where the equation
\begin{equation}
\label{jloc}
 \Lambda(\theta_r) = 0
\end{equation}
is satisfied, with the additional condition that the first
non-zero derivative of $\Lambda(\theta)$ at $\theta_r$ is odd.

Given that under the general conditions specified above the
entropy of entanglement can be expressed as an average over $\UN$,
it is natural to ask whether under different conditions it can be
expressed as an average over random matrices drawn from the other
classical groups.  The question is: how are the symmetries of the
Hamiltonian reflected in the group which determines the entropy of
entanglement?

We begin with the orthogonal group $\ONep$.  The analogue of the
Heine-Szeg\H{o} identity in this case relates group averages to
the determinant of a sum of Toeplitz and Hankel matrices.  A
straightforward calculation generalizing that given above shows
that this can be arranged for the spectral determinant $D_N$ if
(and only if) $\gamma=0$ and $\bA_{jl} = a(j-l) + a(j+l)$
where, because of the periodic boundary conditions, $a$ must be a
function on $\mathbb{Z}/M\mathbb{Z}$ and must also be even in
order for $\bA$ to be symmetric.  Note that the Hamiltonians in
this class are not translation invariant.  The properties of $D_N$
are the same as those in the unitary case, except that $T_N[g]$ is
the sum of a Toeplitz and a Hankel matrix. The symbol has the same
general form as in the unitary case.

The calculations for the other compact groups follow exactly the
same pattern except that for $\SptN$ and $\ONen$, $\bA_{jk} =
a(j-k) - a(j+k +2)$, and for $\ONo{\pm}$, $\bA_{jk} = a(j-k) \mp
a(j+k+1)$.  Again, in these cases the Hamiltonians are not
translation invariant.

The asymptotics of the entropy of entanglement can now be
calculated using the Fisher-Hartwig conjecture for the determinant
$D_N$ in the unitary case~\cite{FH68} and recent generalizations
of this conjecture in the other cases~\cite{BE02, FF04}, and then
by computing the integral in~\eqref{JKidea2}.  The result is that
as $N\rightarrow \infty$
\begin{equation}
\label{mainresult}
 E_{\rm P} \sim \frac{2^{w_{\G}} R}{6}\log_2 N,
\end{equation}
where $R$ is the number of solutions of~\eqref{jloc} in the
interval $[0, \pi)$ and
\begin{equation}
w_{\G} = \begin{cases} 1 & \text{if the average is over $\UN$} \\
                        0 & \text{otherwise.}
          \end{cases}
\end{equation}

The asymptotic relation~\eqref{mainresult} represents our main
result.  In the unitary case, comparing with the results
of~\cite{HLW94, Kor04, CC04}, it provides an  explicit formula for
the central charge, which may be seen to depend in a non-trivial
way on the geometry of the Hamiltonian. In the case of the other
classical compact groups, when the Hamiltonian is not translation
invariant, the conformal-field-theoretic results do not apply
directly.  The factor $2^{w_{{\rm G}}}$ is universal, depending
only on the symmetries determining the classical compact group to
be averaged over.  The factor $R$ is Hamiltonian-dependent.  For
the XX model, which is an example with unitary symmetry, $R=1$
and~\eqref{mainresult} coincides with the formula derived
in~\cite{JK03}.

Lower order terms in the Fisher-Hartwig conjecture and its
generalizations lead directly to general formulae for the
next-to-leading-order (constant) term $\tilde{\kappa}$ in the
asymptotics of the entropy of entanglement when $N \rightarrow
\infty$.  For the unitary group we find
\begin{equation}
\label{O1term} \tilde{\kappa}_{\UN} = \frac{R}{3\ln2}\left( K -
6I_3\ln2\right),
\end{equation}
where $I_3$ is a constant evaluated in~\cite{JK03} to be
$0.0221603\dots$ and
\begin{multline}
K = 1 + \gamma_{{\rm E}} + \frac{1}{R}\sum_{r=1}^R\ln\abs{1 -
\rme^{\rmi 2 \theta_r}} \\ -\frac{2}{R}\sum_{1 \le r < s \le R}
(-1)^{(r + s)} \ln \left |\frac{1 - \rme^{\rmi\left(\theta_r -
\theta_s\right)}}{1 - \rme^{\rmi \left(\theta_r +
\theta_s\right)}}\right|.
\end{multline}
Here $\gamma_{{\rm E}}$ is Euler's constant. When $R=1$ this
equation reduces to the result of Jin and Korepin for the XX
model.  For the other compact groups we find, similarly, that
$\tilde{\kappa}=\tilde{\kappa}_{\UN}/2+R/6$.

We end with some general remarks.  First, the circular ensembles
of random matrices may be seen to play a special role in the
context of the spin chains and boundary conditions we have
considered here. It would be interesting to know whether the other
random matrix ensembles may be used to describe systems with
different interactions and boundary conditions.   Second, we note
that away from critical points $\Lambda(\theta)$ is continuous,
and so $R=0$, which is consistent with previous observations that
the logarithmic growth of $E_{{\rm P}}$ is a critical phenomenon.
Our approach determines the limiting value of $E_{{\rm P}}$ away
from critical points too.  Finally, the calculations and results
described above extend straightforwardly to spin-spin correlations
in the families of quantum spin chains we have considered.

We gratefully acknowledge discussions with Estelle Basor, Peter
Forrester and Noah Linden, and the kind hospitality of the Isaac
Newton Institute, Cambridge.


\begin{thebibliography}{99}
\bibitem{Vira} The central charge of a conformal field theory may be defined (see,
e.g.,~\cite{Kor04}) to be the limit as $z$ tends to zero of
$z^4<T(z)T(0)>$, where $<T(z)T(0)>$ is the autocorrelation
function of the $zz$-component of the energy-momentum tensor
$T_{\mu,\nu}(z)$, $z$ is the complex space-time variable
$z=x+ivt$, and $v$ is the Fermi velocity.  Roughly speaking, the
central charge counts the number of gapless degrees of freedom in
the field theory.
\end{thebibliography}
\end{document}